\documentclass[nofootinbib,twocolumn,showpacs]{revtex4}
\usepackage{graphicx}
\usepackage{amsfonts}
\usepackage{amssymb}
\usepackage{amsmath}
\usepackage{latexsym}

\begin{document}

\title{Classical confinement of test particles in
higher-dimensional models: stability criteria and a new energy
condition}

\author{Sanjeev S.~Seahra}
\email{ssseahra@uwaterloo.ca} \affiliation{Department of Physics,
University of Waterloo, Waterloo, Ontario, N2L 3G1, Canada}

\date{September 8, 2003}

\setlength\arraycolsep{2pt}
\newcommand*{\di}{\partial}
\newcommand*{\V}{{\mathcal V}^{(k)}_d}
\newcommand*{\volume}{\sqrt{\sigma^{(k,d)}}}
\newcommand*{\OneTwo}{{(1,2)}}
\newcommand*{\onetwo}{{1,2}}
\newcommand*{\Lm}{{\mathcal L}_m}
\newcommand*{\stm}{{\textsc{stm}}}
\newcommand*{\Hm}{{\mathcal H}_m}
\newcommand*{\hatHm}{\hat{\mathcal H}_m}
\newcommand*{\Ldust}{{\mathcal L}_\mathrm{dust}}
\newcommand*{\maxsym}{{\mathbb S}_d^{(k)}}
\newcommand*{\ansatz}{{\emph{ansatz}}}
\newcommand*{\ds}[1]{ds^2_\text{\tiny{($#1$)}}}
\newcommand*{\kret}[1]{\mathfrak{K}_\text{\tiny{($#1$)}}}
\newcommand*{\vol}{\mathrm{vol}\,}

\begin{abstract}

We review the circumstances under which test particles can be
localized around a spacetime section $\Sigma_0$ smoothly contained
within a codimension-1 embedding space $M$.  If such a confinement
is possible, $\Sigma_0$ is said to be totally geodesic. Using
three different methods, we derive a stability condition for
trapped test particles in terms of intrinsic geometrical
quantities on $\Sigma_0$ and $M$; namely, confined paths are
stable against perturbations if the gravitational stress-energy
density on $M$ is larger than that on $\Sigma_0$, as measured by
an observed travelling along the unperturbed trajectory.  We
confirm our general result explicitly in two different cases: the
warped-product metric \ansatz~for $(n+1)$-dimensional Einstein
spaces, and a known solution of the 5-dimensional vacuum field
equation embedding certain 4-dimensional cosmologies.  We conclude
by defining a \emph{confinement energy condition} that can be used
to classify geometries incorporating totally geodesic
submanifolds, such as those found in thick braneworld and other
5-dimensional scenarios.

\end{abstract}


\maketitle

\raggedbottom

\section{Introduction}\label{sec:introduction}

The past half-decade has seen a notable upswing in interest in
non-compact higher-dimensional theories of physics.  Most of this
attention can be attributed to recent advances in string theory,
which have postulated that we are living on a $(3+1)$-dimensional
hypersurface embedded within some higher-dimensional manifold.
Such ``braneworld'' scenarios have been extensively analyzed in
the literature, and have been used to address issues such as the
hierarchy problem of particle physics
\cite{Ark98a,Ark98b,Ran99a,Ran99b}, as well as the idea that the
post-inflationary epoch of our universe was preceded by the
collision of D3-branes \cite{Kho01,Buc02}.  In all fairness, it
should be mentioned that the current flurry of interest in
braneworld scenarios has been preceded by numerous other models
making use of large or infinite extra dimensions \cite{Jos62,Aka82,Rub83,%
Vis85,Gib87,Wes92}.

In some braneworld scenarios, the idea of non-compact extra
dimensions is made more palatable by postulating that the
particles and fields of the standard model are confined to the
brane universe.  If we adopt the most conservative point of view,
the notion of confinement is a prerequisite for any reasonable
theory with non-compact extra dimensions; without such an
assumption, the fact that we so not commonly see objects flying
off in unseen directions becomes a thorny issue.  In the context
of a particular string theory-inspired model put forth by Horava
\& Witten \cite{Hor96a,Hor96b}, lower-dimensional confinement is a
natural consequence of the idea that standard model degrees of
freedom are associated with open strings that have endpoints
residing on a D$p$-brane. Conversely, since gravitational degrees
of freedom are associated with closed strings, the graviton in
such models is assumed to propagate both in the bulk and on the
brane. Phenomenological 5-dimensional realizations based this idea
model the brane as a 4-dimensional domain wall or defect
\cite{Ran99a,Ran99b}. The discontinuity in the 5-geometry about
the defect forces the graviton ground state to be sharply
localized on the brane, which allows for the recovery of standard
Newtonian gravitation in the low-energy limit.  This kind of
localization extends to other types of fields, thus representing a
sort of concretization of the confinement mechanism envisioned in
the original string model. In addition, if the matter localized on
the brane satisfies the appropriate energy conditions and the
$\mathbb{Z}_2$ symmetry is obeyed, one can show that test
particles can be gravitationally confined to a small region about
the defect \cite{Sea03a}.  This acts as a classical confinement
mechanism.

A natural generalization of models involving thin geometric
defects is scenarios involving thick, smooth domain walls
\cite{DeW99,Csa00}. There are a couple of reasons to consider such
models:  First, there is a natural minimum length in string theory
given by the string scale, so the idea of an infinitely thin
geometric feature is somewhat suspect even in a phenomenological
model. Second, one would like to see these braneworld scenarios
resulting from some genuine solutions of supergravity, which are
\emph{a priori} smooth and differentiable manifolds.  The question
is: what becomes of the confinement paradigm in bulk manifolds
devoid of thin domain walls?  For test particles in scenarios with
one extra dimension, the answer is well known: if the brane has
vanishing extrinsic curvature, geodesics may be naturally
hypersurface-confined without the invocation of external
non-gravitational forces.  A surface with zero extrinsic curvature
is sometimes called totally geodesic.\footnote{An alternative name
for a totally geodesic submanifold is ``geodesically complete.''}
But what of the stability of the trajectories confined on these
surfaces? That is, if one perturbs a confined trajectory slightly
off of a totally geodesic submanifold, will it naturally return or
not?  In other words, under which conditions is a totally geodesic
hypersurface gravitationally attractive? For obvious reasons, such
questions are of direct relevance to any serious attempt to
classically describe our universe as a smoothly-embedded
hypersurface on which we are gravitationally trapped.  It is
possible that this classical stability issue is irrelevant at the
quantum level --- perhaps because stable particle confinement can
be guaranteed by other means --- but for the purposes of this
study we will assume that the classical formalism is applicable.

In this paper, we propose to address these issues $n$-dimensional
totally geodesic submanifolds smoothly embedded in a space of one
higher dimension, with either timelike or spacelike signature. We
will utilize quite general methods that will ensure our results
apply to any geometry and choice of coordinates in the bulk or on
the submanifold.  In Section \ref{sec:geometry}, we describe our
geometric construction.  In Section \ref{sec:confinement}, we
review the covariant splitting of test particle equations of
motion developed in \cite{Sea02,Sea03a,Sea03b} and use it to
derive the zero-extrinsic curvature condition for totally geodesic
submanifolds.  Then, we find the stability condition for the
confined trajectories, which is that the double contraction of the
particle's velocity with the Ricci tensor of the bulk is greater
than the double contraction with the Ricci tensor of the
submanifold.  In more physical terms, the stability of trapped
particles demands that the locally measured gravitational density
of the bulk is bigger than the density of the effective
lower-dimensional matter living on the brane.\footnote{Roughly
speaking, the gravitational density of a given matter-energy
distribution differs from the ordinary density by terms involving
the pressure.  For example, according to an observer comoving with
a $(n+1)$-dimensional perfect fluid, the gravitational density ---
as we define it below --- is $[(n-2)\rho + np]/(n-1)$.  It is
important because it appears naturally in the Raychaudhuri
equation, as we will see in Section \ref{sec:raychaudhuri}.} We
briefly discuss the nature of the latter, emphasizing that the
stress-energy content of the submanifold --- as perceived by an
observer ignorant of an extra dimension --- is made up from
contributions from the ``real'' higher-dimensional matter as well
as the bulk Weyl tensor.  For good measure, we derive the
stability condition using two additional methods: the geodesic
deviation equation in Section \ref{sec:deviation}, and the
Raychaudhuri equation in Section \ref{sec:raychaudhuri}. We
confirm the correctness of our general result for the special case
of the warped-product metric \ansatz~in Section \ref{sec:warped},
and consider a simple 5-dimensional model of the solar
neighborhood.  In Section \ref{sec:lmw}, we show that our
stability condition is also correct in the Liu-Mashoon-Wesson
solution \cite{Liu95,Liu01} of the 5-dimensional vacuum field
equations. Section \ref{sec:summary} summarizes our work and
presents the \emph{confinement energy condition}, which ensures
that all timelike trajectories on a totally geodesic submanifold
in a given bulk geometry will be stable.  This energy condition
can be used to classify solutions of the thick braneworld on other
5-dimensional scenarios.

\paragraph*{Conventions.}  Uppercase Latin indices run from 0 to
$n$, while lowercase Greek indices run from 0 to $n-1$.
Higher-dimensional curvature tensors are distinguished from their
lower-dimensional counterparts by hats.  Higher and lower
dimensional covariant differentiation operators are denoted by
$\nabla_A$ or $\nabla_\alpha$, respectively.  A center dot
indicates the scalar product between higher-dimensional vector
fields; i.e., $u \cdot v \equiv u^A v_A$.

\section{Geometric construction}\label{sec:geometry}

We will be concerned with an $(n+1)$-dimensional manifold
$(M,g_{AB})$ on which we place a coordinate system $x \equiv \{
x^A \}$.  Sometimes, we will refer to $M$ as the ``bulk
manifold.''  In our working, we will allow for two possibilities:
either there is one timelike and $n$ spacelike directions tangent
to $M$, \emph {or} there are two timelike and $(n-1)$ spacelike
directions tangent to $M$. Hence, the signature of $g_{AB}$ is
\begin{equation}
    \mathrm{sig}(g_{AB}) = (-+\cdots+\varepsilon),
\end{equation}
where $\varepsilon = \pm 1$.  We introduce a scalar function
\begin{equation}
    \ell = \ell(x),
\end{equation}
which defines our foliation of the higher-dimensional manifold
with the hypersurfaces given by $\ell$ = constant, denoted by
$\Sigma_\ell$.  If there is only one timelike direction tangent to
$M$, we assume that the vector field $n^A$ normal to $\Sigma_\ell$
is spacelike.  If there are two timelike directions, we take the
unit normal to be timelike.  In either case, the submanifold
tangent to a given $\Sigma_\ell$ hypersurface contains one
timelike and $(n-1)$ spacelike directions; that is, each
$\Sigma_\ell$ hypersurface corresponds to an $n$-dimensional
Lorentzian spacetime.  The normal vector to the $\Sigma_\ell$
slicing is given by
\begin{equation}\label{2:normal definition}
    n_A = \varepsilon \Phi \, \di_A \ell, \quad n \cdot n = \varepsilon.
\end{equation}
The scalar $\Phi$ which normalizes $n^A$ is known as the lapse
function.  We define the projection tensor as
\begin{equation}
\label{2:induced metric}
    h_{AB} = g_{AB} - \varepsilon n_A n_B.
\end{equation}
This tensor is symmetric ($h_{AB} = h_{BA}$) and orthogonal to
$n_A$.  We place an $n$-dimensional coordinate system on each of
the $\Sigma_\ell$ hypersurfaces $y \equiv \{ y^\alpha \}$.  The
$n$ holonomic basis vectors
\begin{equation}
    e^A_\alpha = \frac{\di x^A}{\di y^\alpha}, \quad n \cdot
    e_\alpha = 0
\end{equation}
are by definition tangent to the $\Sigma_\ell$ hypersurfaces and
orthogonal to $n^A$.  It is easy to see that $e^A_\alpha$ behaves
as a vector under coordinate transformations on $M$ [$\phi:x
\rightarrow \bar{x} (x)$] and a one-form under coordinate
transformations on $\Sigma_\ell$ [$\psi:y \rightarrow
\bar{y}(y)$].  We can use these basis vectors to project
higher-dimensional objects onto $\Sigma_\ell$ hypersurfaces.  For
example, for an arbitrary one-form on $M$ we have
\begin{equation}
    T_\alpha = e_\alpha^A T_A = e_\alpha \cdot T.
\end{equation}
Here $T_\alpha$ is said to be the projection of $T_A$ onto
$\Sigma_\ell$.  Clearly $T_\alpha$ behaves as a scalar under
$\phi$ and a one-form under $\psi$.  The induced metric on the
$\Sigma_\ell$ hypersurfaces is given by
\begin{equation}
    h_{\alpha\beta} = e^A_\alpha e^B_\beta g_{AB} = e^A_\alpha
    e^B_\beta h_{AB}.
\end{equation}
Just like $g_{AB}$, the induced metric has an inverse:
\begin{equation}
    h^{\alpha\gamma} h_{\gamma\beta} = \delta^{\alpha}{}_{\beta}.
\end{equation}
The induced metric and its inverse can be used to raise and lower
the indices of tensors tangent to $\Sigma_\ell$, and change the
position of the spacetime index of the $e^A_\alpha$ basis vectors.
This implies
\begin{equation}
    e_A^\alpha e^A_\beta = \delta^\alpha{}_\beta.
\end{equation}
Also note that since $h_{AB}$ is entirely orthogonal to $n^A$, we
can express it as
\begin{equation}\label{2:induced decomposition}
    h_{AB} = h_{\alpha\beta} e^\alpha_A e^\beta_B.
\end{equation}
At this juncture, it is convenient to introduce our definition of
the extrinsic curvature $K_{\alpha\beta}$ of the $\Sigma_\ell$
hypersurfaces:\index{extrinsic curvature}
\begin{equation}\label{2:extrinsic def}
    K_{\alpha\beta} = e^A_\alpha e^B_\beta \nabla_A n_B =
    \tfrac{1}{2} e^A_\alpha e^B_\beta \hat\pounds_n h_{AB}.
\end{equation}
Note that the extrinsic curvature is symmetric ($K_{\alpha\beta} =
K_{\beta\alpha}$).  It may be thought of as the derivative of the
induced metric in the normal direction.  This $n$-tensor will
appear often in what follows.

We will also require an expression that relates the
higher-dimensional covariant derivative of $(n+1)$-tensors to the
lower-dimensional covariant derivative of the corresponding
$n$-tensors.  We have that the $n$-dimensional Christoffel symbols
are given by
\begin{equation}\label{2:Christoffel}
    \Gamma^\alpha_{\beta\gamma} = e_\gamma^B e_A^\alpha \nabla_B e^A_\beta.
\end{equation}
This allows us to deduce that for one-forms, the following
relation holds:
\begin{equation}
    \nabla_\beta T_\alpha = e^B_\beta e^A_\alpha \nabla_B
    (h_A{}^C T_C),
\end{equation}
where $\nabla_B$ is the covariant derivative on $M$ defined with
respect to $g_{AB}$ and $\nabla_\beta$ is the covariant derivative
on $\Sigma_\ell$ defined with respect to $h_{\alpha\beta}$.  The
generalization to tensors of higher rank is obvious.  It is not
difficult to confirm that this definition of $\nabla_\alpha$
satisfies all the usual requirements imposed on the covariant
derivative operator.

Finally, we note that $\{ y, \ell \}$ defines an alternative
coordinate system to $x$ on $M$.  The appropriate diffeomorphism
is
\begin{equation}\label{2:diffeo}
    dx^A = e_\alpha^A dy^\alpha + \ell^A d\ell,
\end{equation}
where
\begin{equation}\label{2:flow of l vector}
    \ell^A =  \left( \frac{\di x^A}{\di \ell} \right)_{y^\alpha =
    \mathrm{const.}}
\end{equation}
is the vector tangent to lines of constant $y^\alpha$.  We can
always decompose higher dimensional vectors into the sum of a part
tangent to $\Sigma_\ell$ and a part normal to $\Sigma_\ell$.  For
$\ell^A$ we write
\begin{equation}\label{2:l vector}
    \ell^A = N^\alpha e_\alpha^A + \Phi n^A.
\end{equation}
This is consistent with $\ell^A \di_A \ell = 1$, which is required
by the definition of $\ell^A$, and the definition of $n^A$.  The
$n$-vector $N^\alpha$ is the shift vector, which describes how the
$y^\alpha$ coordinate system changes as one moves from a given
$\Sigma_\ell$ hypersurface to another.  Using our formulae for
$dx^A$ and $\ell^A$, we can write the higher dimensional line
element as
\begin{eqnarray}\nonumber
    \ds{M} & = & g_{AB} \, dx^A dx^B \\ \nonumber & = & h_{\alpha\beta}
    (dy^\alpha + N^\alpha d\ell) (dy^\beta + N^\beta d\ell) \\ & &
    + \varepsilon \Phi^2 d\ell^2.\label{2:lapse and shift line
    element}
\end{eqnarray}
This reduces to $\ds{\Sigma_\ell} = h_{\alpha\beta} dy^\alpha
dy^\beta$ if $d\ell = 0$.  It is also possible to express the
extrinsic curvature in terms of $\Phi$ and $N^\alpha$:
\begin{equation}\label{2:alternate extrinsic}
    K_{\alpha\beta} = \frac{1}{2\Phi} ( \di_\ell - \pounds_N )
    h_{\alpha\beta},
\end{equation}
where $\pounds_N$ is the Lie derivative in the direction of the
shift vector.

In this paper, we will be primarily concerned with the
Gaussian-normal coordinate gauge that has been termed canonical by
some authors \cite{Mas94}.  This is defined by the following
choices of foliation parameters:
\begin{equation}\label{canonical}
    \Phi = 1, \quad N^\alpha = 0.
\end{equation}
Obviously, this choice will result in significant simplification
of many of the preceding and following formulae.

\section{Confinement of test particles}\label{sec:confinement}

The equations of motion for a test particle travelling through $M$
are taken to be
\begin{equation}
    u^A \nabla_A u^B = \mathcal{F}^B, \quad u \cdot u = \kappa,
    \quad u^A = \frac{dx^A}{d\lambda},
\end{equation}
where $\kappa = -1, 0, +1$ to allow for massive, null and
tachyonic particles respectively, $\lambda$ is an affine
parameter, and $\mathcal{F}$ is some non-gravitational force per
unit mass.  One can decompose these equations into relations
involving the particle's velocity tangent to the $\Sigma_\ell$
foliation $u^\alpha = e^\alpha \cdot u$ and parallel to the normal
direction $u_n = n \cdot u$.  This was first done in \cite{Sea02}
for a 5-dimensional model with a spacelike extra dimension and
pure geodesic motion, then generalized to accelerated trajectories
and an extra dimension of arbitrary signature in \cite{Sea03a},
and further adapted to arbitrary dimension and refined notation in
\cite{Sea03b}. Here, we will merely adopt the final results, which
are:
\begin{subequations}\label{2:split equations}
\begin{eqnarray}\nonumber
    u^\alpha \nabla_\alpha u^\beta & = & \varepsilon u_n [
    u_n \di^\beta \ln\Phi -2 K^{\alpha\beta} u_\alpha - \\ & &
    \Phi^{-1} (\di_\ell - \pounds_N)u^\beta ] + \mathcal{F}^\beta,
    \label{2:n-dimensional part} \\ \label{2:extra part}
    \dot{u}_n & = & K_{\alpha\beta} u^\alpha u^\beta - u_n u^\alpha
    \di_\alpha \ln\Phi
    + \mathcal{F}_n, \\ \label{2:split norm}
    \kappa & = & h_{\alpha\beta} u^\alpha u^\beta + \varepsilon u_n^2.
\end{eqnarray}
\end{subequations}
where we have defined $\mathcal{F}^\alpha \equiv e^\alpha \cdot
\mathcal{F}$, $\mathcal{F}_n \equiv n \cdot \mathcal{F}$, and an
overdot indicates $d/d\lambda$.  We can express both $u^\alpha$
and $u_n$ in terms of the foliation parameters:
\begin{equation}
    u^\alpha = \dot{y}^\alpha + \dot\ell N^\alpha, \quad u_n =
    \varepsilon \Phi \dot\ell.
\end{equation}
This form of the equations of motion has the virtue of being
written entirely in terms of tensorial quantities on
$\Sigma_\ell$, which makes it invariant under $n$-dimensional
coordinate transformations.  Also note that one of the equations
(\ref{2:split equations}) is redundant; for example, if one
contracts (\ref{2:n-dimensional part}) with $u_\beta$ and makes
use of (\ref{2:split norm}), one can recover (\ref{2:extra part}).

Now, if a test particle is confined to a given $\Sigma_\ell$
hypersurface, its $\ell$-coordinate must obviously be constant.
This implies $u_n \equiv 0$, which by (\ref{2:extra part}) yields
\begin{equation}
    0 = K_{\alpha\beta} u^\alpha u^\beta + \mathcal{F}_n.
\end{equation}
In other words, if the normal force per unit mass equals
$-K_{\alpha\beta} u^\alpha u^\beta$, then the particle can be
hypersurface-confined.  Since this quantity is quadratic in the
particle's $n$-velocity, we can identify it as the generalized
centripetal acceleration in curved space.  Indeed, in \cite{Sea02}
it was shown that when a particle is confined to the world tube of
a circle $\mathbb{R} \times S$ embedded in 3-dimensional Minkowski
space, $\mathcal{F}_n$ reduces to the familiar $v^2/r$ from
elementary mechanics.

Now, if one member $\Sigma_0$ of the $\Sigma_\ell$ foliation
satisfies $K_{\alpha\beta} = 0$, then it is obvious that no
external centripetal force $\mathcal{F}^A$ is required to ensure
confinement. Indeed, when the extrinsic curvature vanishes one
solution of the freely-falling equations of motion is
\begin{equation}
    \dot{y}^\alpha \nabla_\alpha \dot{y}^\beta = 0, \quad \dot\ell
    = 0;
\end{equation}
i.e., the geodesics of $\Sigma_0$ are also geodesics of $M$. As
mentioned in Section \ref{sec:introduction}, surfaces with this
property are termed totally geodesic and they represent
equilibrium surfaces for freely-falling test particles.  We want
to know how to tell if such surfaces represent stable or unstable
equilibria.

To answer this, we can attempt to linearize the equations of
motion about $\Sigma_0$; that is, we consider the motion of a test
particle very close to the equilibrium hypersurface.  To simplify
matters, we will adopt the canonical gauge (\ref{canonical})
discussed above. Then, it is straightforward to derive expressions
for $\di_\ell h_{\alpha\beta}$ and $\di_\ell K_{\alpha\beta}$
\cite{Sea03b}:
\begin{subequations}\label{3:evolution}
\begin{eqnarray} \label{3:evolution 1}
    \di_\ell h_{\alpha\beta} & = & 2 K_{\alpha\beta},
    \\ \label{3:evolution 2}
    \di_\ell K_{\alpha\beta} & = & K_{\alpha}{}^\mu
    K_{\mu\beta} -E_{\alpha\beta},
\end{eqnarray}
\end{subequations}
where $E_{\alpha\beta} \equiv e^A_\alpha e^B_\beta n^C n^D
\hat{R}_{ACBD}$.  This $n$-tensor can be related back to more
familiar quantities by suitable manipulation of the Gauss-Codazzi
relations:
\begin{equation}\label{GC eqn}
    \hat{R}_{AB} e^A_\alpha e^B_\beta = R_{\alpha\beta} + \varepsilon
    [ E_{\alpha\beta} + K_{\alpha}{}^{\mu} ( K_{\beta\mu} -
    K h_{\beta\mu} ) ].
\end{equation}
Now, without loss of generality we can suppose that $\Sigma_0$
corresponds to $\ell = 0$; from this, it follows that $|\ell|$ is
``small'' in our approximation.\footnote{Technically, we require
that $\ell$ be small compared to the characteristic size of the
components of the curvature tensors on $\Sigma_0$ and $M$ in order
to justify dropping the $\mathcal{O}(\ell^2)$ terms in
(\ref{extrinsic approx}).  In practical terms, this means $\ell$
should be much less than the radii of curvature of both
manifolds.}  If we let ${}^0 h_{\alpha\beta}$ denote the induced
metric on $\Sigma_0$, ${}^0 R_{\alpha\beta}$ the Ricci-tensor on
$\Sigma_0$, etc.; then we have
\begin{subequations}\label{extrinsic approx}
\begin{eqnarray}\label{induced derivative}
    h_{\alpha\beta} & = & {}^0 h_{\alpha\beta} +
    \mathcal{O}(\ell^2), \\
    K_{\alpha\beta} & = & \varepsilon ( {}^0 R_{\alpha\beta} - {}^0
    \hat{R}_{AB} e^A_\alpha
    e^B_\beta ) \ell +  \mathcal{O}(\ell^2).
\end{eqnarray}
\end{subequations}
Furthermore, we suppose that the $n$-velocity of our particle to
be approximately tangent to a geodesic on $\Sigma_0$:
\begin{equation}
    u^\alpha = U^\alpha + \delta u^\alpha, \quad
    U^\alpha \, \nabla_\alpha \, U^\beta = 0.
\end{equation}
Here, $\delta u^\alpha$ is considered to be a small quantity; that
is, we are really considering perturbations of the confined
trajectory tangent to $ U^A = U^\alpha e^A_\alpha$.  (We will
comment on the validity of this assumption below.)  Then, to
lowest order in small quantities, equation (\ref{2:extra part})
reduces to
\begin{equation}\label{linear extra}
    \ddot\ell = ( {}^0 R_{\alpha\beta} - {}^0 \hat{R}_{AB} e^A_\alpha
    e^B_\beta ) U^\alpha U^\beta \ell + \cdots
\end{equation}
If $\Sigma_0$ is a stable equilibrium for test particles, we
require that $\ddot\ell/\ell < 0$.  This condition translates into
the following condition for the confinement of test particles on
$\Sigma_0$:
\begin{equation}
    ( {}^0 R_{\alpha\beta} - {}^0 \hat{R}_{AB} e^A_\alpha
    e^B_\beta ) U^\alpha U^\beta < 0.
\end{equation}

In order to interpret this result, we define the following
quantities:
\begin{subequations}\label{density defs}
\begin{eqnarray}
    \rho_g^{(n+1)} & \equiv & {}^0 \hat{R}_{AB} e^A_\alpha
    e^B_\beta U^\alpha U^\beta, \\
    \rho_g^{(n)} & \equiv & {}^0 R_{\alpha\beta} U^\alpha U^\beta.
\end{eqnarray}
\end{subequations}
Here, $\rho_g^{(n+1)}$ is our definition of the local
gravitational density of higher-dimensional --- or real --- matter
as measured by an observer freely-falling along $\Sigma_0$.  It is
guaranteed to be positive if the strong energy condition is
satisfied in the bulk, or at least in the vicinity of the totally
geodesic surface.  Note that it is possible to define the
gravitational density with a different normalization constant to
obtain a ``nicer'' expression in the perfect fluid case.  That is,
for a $(3+1)$-dimensional perfect fluid, an observer comoving with
the fluid will measure $\rho_g^{(4)} = \tfrac{1}{2}(\rho + 3p)$
using our definition.  One might be motivated to give a different
definition of $\rho_g^{(4)}$ that does not involve the
$\tfrac{1}{2}$ prefactor, but such a choice would unnecessarily
complicate the following discussion with spurious
dimension-dependant terms.

Now, it is clear that $\rho_g^{(n)}$ is the lower-dimensional
cousin of $\rho_g^{(n+1)}$, but the precise interpretation is a
little more subtle.  Imagine an observer living on $\Sigma_0$ that
is entirely ignorant of the $\ell$-direction.  Assuming that this
observer can measure the local $n$-geometry and believes in the
Einstein field equations, he will interpret the Einstein
$n$-tensor of $\Sigma_0$ as being proportional to some effective
stress-energy tensor.  In other words, he will conclude that his
local $n$-geometry is determined by an effective distribution of
matter-energy.  From this, it follows that $\rho_g^{(n)}$ is the
gravitational density of the effective $n$-dimensional matter.

A natural question is: how is the lower-dimensional effective
matter related to the ``real'' higher-dimensional distribution?
It is not hard to derive the following expression for the
stress-energy tensor on $\Sigma_0$ from the Gauss-Codazzi
relations:
\begin{eqnarray}\nonumber
    G^{\alpha\beta} & = & - \varepsilon E^{\alpha\beta} + \hat{G}^{AB}
    e_A^\alpha e_B^\beta \\
    & & - \left[
    \hat{G}_{AB} h^{AB} - \left( \tfrac{n-2}{n-1}
    \right) \mathrm{Tr}( \hat{G} ) \right] h^{\alpha\beta}
     \label{3:general einstein},
\end{eqnarray}
where all quantities are understood to be evaluated on $\Sigma_0$
and we have made note of $K_{\alpha\beta} = 0$. The last two terms
on the right are explicitly related to the higher-dimensional
stress-energy tensor, but the first can be re-written using
\begin{eqnarray}\nonumber
    E_{\alpha\beta} & = & e^A_\alpha e^B_\beta n^C  n^D \hat{C}_{ACBD}
    + \tfrac{1}{n(n-1)} \varepsilon \hat{R} h_{\alpha\beta} \\ & & +
    \tfrac{1}{n-1} \left( \hat{R}_{AB} n^A n^B h_{\alpha\beta} +
    \varepsilon \hat{R}_{AB} e^A_\alpha e^B_\beta \right) .
\end{eqnarray}
The first term on the right shows how $E_{\alpha\beta}$ is in part
determined by the Weyl $(n+1)$-tensor $\hat{C}_{ABCD}$ on $M$.
This curvature tensor is related to the geometrical or
gravitational degrees of freedom of the bulk, which are not
directly fixed by the $(n+1)$-dimensional field equations.  This
implies that the Einstein $n$-tensor $G_{\alpha\beta}$ on $M$ is
not entirely determined by the distribution of higher-dimensional
stress-energy --- there is a purely geometric contribution from
the appropriate projection of $\hat{C}_{ABCD}$.  We call this
contribution the \emph{induced} \cite{Wes99} or \emph{shadow}
\cite{Fro03} matter stress-energy tensor because it represents a
source of the lower-dimensional Einstein equation that cannot be
unambiguously attributed to any ``real'' higher-dimensional
fields. It follows that the $n$-dimensional effective
gravitational density contains contributions from both real and
induced matter.

The final point we wish to discuss in this section has to do with
the validity of our approximations.  Recall that above, in order
to derive equation (\ref{linear extra}), we assumed that $\delta
u^\alpha$ was a small quantity.  We now want to describe under
which circumstances this hypothesis holds by substituting the
expansion $u^\alpha = U^\alpha + \delta u^\alpha$ in equation
(\ref{2:extra part}) in the canonical gauge with $\mathcal{F} =
0$. For this purpose it is useful to assume that $U^\alpha =
U^\alpha(y(\lambda))$, or equivalently $\di_\ell U^\alpha = 0$.
Coupled with equation (\ref{induced derivative}), this implies
that $U^\alpha \nabla_\alpha U^\beta = \mathcal{O}(\ell^2)$. Under
such circumstances, we obtain
\begin{eqnarray}\nonumber
    \frac{d}{d\lambda} \delta u^\alpha & = &  {}^{0}E^{\alpha\beta}
    (U_\beta - \delta u_\beta) \frac{d}{d\lambda} \ell^2 -
    \Gamma_{\beta\gamma}^\alpha \, \delta u^\beta \, \delta
    u^\gamma \\ & & - \delta u^\beta \, \nabla_\beta U^\alpha,
    \label{velocity perturbation}
\end{eqnarray}
where we have made use of $K_{\alpha\beta} \approx -{}^0
E_{\alpha\beta} \times \ell$.  It is clear that in order to have
$\delta u^\alpha$ be consistently ``small'', we must have that the
only term on the right inhomogeneous in $\delta u^\alpha$ is
negligible. Now, since ${}^0 E_{\alpha\beta} = \varepsilon ( {}^0
\hat{R}_{AB} e^A_\alpha e^B_\beta - {}^0 R_{\alpha\beta} )$ we see
that its components are of the order of the inverse squares of the
curvature lengths of $M$ and $\Sigma_0$.  Then, for our
approximations to be valid, we need that $d(\ell^2)/d\lambda$ be
small compared to the characteristic curvature of $M$ or
$\Sigma_0$, whichever is smaller.  This is a sensible intuitive
bound --- if either the higher- or lower-dimensional manifolds are
highly curved we expect that true confinement will be difficult to
achieve.  So, in addition to demanding that $\ell$ is small in
order to justify $K_{\alpha\beta} \approx -{}^0 E_{\alpha\beta}
\times \ell$, we also need the extra dimensional velocity to be
relatively tiny.

To summarize, we have seen that any $n$-surface $\Sigma_0$
embedded in $M$ is an equilibrium position for freely-falling test
particles if it has vanishing extrinsic curvature. Every geodesic
on $\Sigma_0$ is automatically a geodesic of $M$, hence the
hypersurface is called totally geodesic.  Furthermore, if a given
trajectory confined to $\Sigma_0$ is perturbed off of the
equilibrium surface, the acceleration of the test particle is
towards $\ell = 0$ provided that
\begin{enumerate}

\item[\emph{i})]{the gravitational density of the
higher-dimensional matter is greater than the gravitational
density of the effective lower-dimensional matter on $\Sigma_0$,
as measured by an observer travelling on the unperturbed
trajectory; and,}

\item[\emph{ii})]{both $\ell$ and $d(\ell^2)/d\lambda$ are small
compared with the characteristic curvature scales of $M$ and
$\Sigma_0$.}

\end{enumerate}
Notice that these comments apply to a particular trajectory only.

\section{Two alternative derivations of the stability condition}

The stability condition for freely-falling test particle
trajectories derived above depended on our decomposition of the
higher-dimensional equation of motion (\ref{2:split equations})
and the canonical gauge for the foliation parameters
(\ref{canonical}).  It is possible to derive the same condition
using two different methods that relax one or both of these
assumptions, which is what we do in this section.

\subsection{From the geodesic deviation
equation}\label{sec:deviation}

Consider a freely-falling test particle on $\Sigma_0$ that has an
$(n+1)$-velocity $U^A = e^A_\alpha U^\alpha$ at a particular
instant of time. As before, we assume that $\Sigma_0$ is totally
geodesic, which means that the test particle will remain confined
on the submanifold for all future times in the absence of
non-gravitational influences.  Now, consider an additional test
particle that is separated from the first object by a vector
$\xi^A = \ell n^A$. Here, $\ell$ is the proper distance separating
the two particles and is assumed to be small. Then, the equation
of geodesic deviation says that
\begin{equation}
    a^A = -R^A{}_{BCD} U^B \xi^C U^D,
\end{equation}
where $a^A$ is the acceleration of $\xi^A$, defined as
\begin{equation}
    a^A = (U \cdot \nabla)^2 \xi^A.
\end{equation}
Now, consider $n \cdot a$.  Making use of $\xi^A = \ell n^A$, we
find the following expression for this scalar product:
\begin{equation}
    n \cdot a = \varepsilon \ddot\ell - \ell h^{AB} (U^C\nabla_C
    n_A)(U^D \nabla_D n_B).
\end{equation}
Here, we have used $g_{AB} = h_{AB} + \varepsilon n_A n_B$, $n^A
U^B \nabla_B n_A = 0$, and $\ddot \ell = (U \cdot \nabla)^2 \ell$.
Because $U^A$ is parallel to $\Sigma_0$, the second term on the
right reduces to $-\ell K_{\alpha\beta} K^{\alpha\gamma} U^\beta
U_\gamma = 0$. Hence, we have
\begin{equation}
    \ddot\ell = -\varepsilon R_{ABCD} n^A U^B n^C U^D \ell =
    -\varepsilon E_{\alpha\beta} U^\alpha U^\beta \ell.
\end{equation}
If we now use (\ref{GC eqn}) to substitute for $E_{\alpha\beta}$,
we immediately recover our previous result (\ref{linear extra}).
Hence, the test particle located just off $\Sigma_0$ will be
accelerated towards $\ell = 0$ if the stability condition from the
previous section $\rho_g^{(n+1)} > \rho_g^{(n)}$ holds.  Notice
that we did not assume the canonical gauge for this derivation.

\subsection{From the Raychaudhuri equation}\label{sec:raychaudhuri}

To establish the stability condition from the Raychaudhuri
equation, we again consider a freely-falling test particle on
$\Sigma_0$ with a trajectory $\gamma$ tangent to $U^A = e_\alpha^A
U^\alpha$. Consider some small $(n-1)$-dimensional region $\delta
V_{n-1} \subset \Sigma_0$ orthogonal to $\gamma$ at some given
time $\lambda_0$.  We extend $\delta V_{n-1}$ a small distance of
$\ell$ on either side of $\Sigma_0$ to define an $n$-dimensional
region $\delta V_n$.  Since $U^A$ is tangent to $\Sigma_0$, our
test particle's trajectory is also orthogonal to $\delta V_n$. Now
consider a geodesic congruence centered about $\gamma$ and
threading every point within $\delta V_n$.  At the moment of
interest, we can take each member of the $(n+1)$-congruence to be
parallel to $\gamma$.  This means that the subset of the total
congruence situated on $\Sigma_0$ is an $n$-dimensional geodesic
congruence on the submanifold.  To evolve the orthogonal regions
forward in time, we imagine that each point in $\delta V_n$ is
glued to the associated geodesic, so the small region deforms in
the same manner as the congruence.  Since the congruence is
instantaneously parallel at $\lambda_0$, we have that
$\dot{\ell}(\lambda_0) = 0$.

Now in the canonical gauge, the volume of $\delta V_n$ is related
to the volume of $\delta V_{n-1}$ by
\begin{equation}
    \vol \delta V_n = 2 \ell \vol \delta V_{n-1}.
\end{equation}
We can define expansion scalars for both the higher- and
lower-dimensional congruences:
\begin{subequations}
\begin{eqnarray}
    \theta_n & = & \frac{d}{d\lambda} \ln (\vol \delta
    V_n) = \nabla_A U^A, \\ \theta_{n-1} & = & \frac{d}{d\lambda} \ln
    (\vol \delta V_{n-1}) = \nabla_\alpha U^\alpha.
\end{eqnarray}
\end{subequations}
It is easy to see that at $\lambda = \lambda_0$ we have
\begin{equation}
    \dot\theta_n = \ddot\ell / \ell + \dot\theta_{n-1},
\end{equation}
where we have made use of $\dot\ell(\lambda_0) = 0$.  The
Raychaudhuri equation applied to each congruence gives that
\begin{subequations}
\begin{eqnarray}
    \dot\theta_n & = & -\nabla^A U^B \nabla_B U_A - {}^0 \hat{R}_{AB} U^A
    U^B,\\
    \dot\theta_{n-1} & = & -\nabla^\alpha U^\beta \nabla_\beta U_\alpha -
    {}^0 R_{\alpha\beta} U^\alpha U^\beta.
\end{eqnarray}
\end{subequations}
A reasonably quick calculation reveals
\begin{eqnarray}
    \nonumber & & \nabla^A U^B \nabla_B U_A \\ \nonumber
    & = & (h^{AC} + \varepsilon n^A n^C) (h^{BD} + \varepsilon n^B
    n^D) \nabla_C U_D \nabla_B U_A \\ \nonumber & = & \nabla^\alpha U^\beta \nabla_\beta
    U_\alpha - 2\varepsilon K_{\alpha\beta} U^\beta e_A^\alpha n^B \nabla_B
    U^A \\ \nonumber & & + (U^\alpha \di_\alpha \ln\Phi)^2 \\ & = & \nabla^\alpha U^\beta \nabla_\beta
    U_\alpha.
\end{eqnarray}
In going from the third to fourth line we have made use of the
fact that $\Sigma_0$ is totally geodesic ($K_{\alpha\beta} = 0$)
and our gauge choice ($\Phi = 1$).  Putting it all together, we
get
\begin{equation}\label{Ray EOM}
    \ddot\ell = \ell[\rho_g^{(n)} - \rho_g^{(n+1)}].
\end{equation}
The geodesics of the congruence will accelerate towards the
equilibrium hypersurface if the quantity in the square brackets is
negative.  This yields the same stability condition as before:
$\rho_g^{(n)} < \rho_g^{(n+1)}$.

\section{Examples}\label{sec:examples}

We have now established the stability condition $\rho_g^{(n)} <
\rho_g^{(n+1)}$ using three different methods in general geometric
backgrounds.  In this section, we give a few examples of specific
manifolds containing totally geodesic hypersurfaces.  For each
case, we explicitly confirm that our stability condition for
confined trajectories  is correct.

\subsection{Warped-product spaces}\label{sec:warped}

For our first example of test particle confinement, we consider
the so-called warped product metric \emph{ansatz}:
\begin{equation}
    \ds{M} = e^\Omega q_{\alpha\beta} dy^\alpha dy^\beta +
    \varepsilon d\ell^2,
\end{equation}
where the warp factor $\Omega = \Omega(\ell)$ is independent of
the $n$-dimensional $y$ coordinates and the warp metric
$q_{\alpha\beta} = q_{\alpha\beta}(y)$ is independent of the extra
dimensional $\ell$.  The induced metric on $\Sigma_\ell$
hypersurfaces is $h_{\alpha\beta} = e^\Omega q_{\alpha\beta}$.  We
will assume that the bulk is an Einstein space satisfying
\begin{equation}
    \hat{G}_{AB} = -\Lambda g_{AB}, \quad \hat{R}_{AB} = \frac{2\Lambda}{n-1} g_{AB}.
\end{equation}
Then, solutions for the warp factor and warp metric are easily
found \cite{Sea03b}:
\begin{subequations}
\begin{eqnarray}\label{warp factor}
    e^{\Omega/2} & = & A
    \begin{cases}
        \cos \omega\ell, & \varepsilon \Lambda > 0, \\
        \cosh \omega\ell, & \varepsilon \Lambda < 0,
    \end{cases} \\ \label{warp Ricci}
    R_{\alpha\beta} & = & \frac{2\Lambda A^2}{n} q_{\alpha\beta},
    \\ \omega^2 & \equiv & \frac{2|\Lambda|}{n(n-1)},
\end{eqnarray}
\end{subequations}
where $A$ is a constant and $R_{\alpha\beta}$ is the Ricci tensor
formed from either the induced or warp metrics (both $n$-tensors
give the same result).  Essentially, the above states that the
warp metric can be taken to be \emph{any} $n$-dimensional solution
of the Einstein field equations sourced by a cosmological constant
$\Lambda_n = A^2 \Lambda (n-2)/n$.

Now, for both of the cases shown in (\ref{warp factor}), it is
clear that the $\ell = 0$ hypersurface $\Sigma_0$ is totally
geodesic.  If we consider some timelike geodesic confined to
$\Sigma_0$, the higher- and lower-dimensional gravitational
densities measured by an observer travelling along that geodesic
are:
\begin{equation}
    \rho_g^{(n+1)} = -\frac{2\Lambda}{n-1}, \quad \rho_g^{(n)} =
    -\frac{2\Lambda}{n},
\end{equation}
where we have used
\begin{equation}
    -1 = U \cdot U = h_{\alpha\beta} U^\alpha U^\beta = A^2 q_{\alpha\beta} U^\alpha
    U^\beta.
\end{equation}
So, the condition that trajectories on $\Sigma_0$ be stable
against perturbations in this case simply reads:
\begin{equation}
    \Lambda < 0;
\end{equation}
i.e., the bulk must have anti-deSitter
characteristics.\footnote{It is a misnomer to call the bulk
anti-deSitter space in this case, it merely satisfies the same
Einstein equations as anti-deSitter space.}

This scenario is simple enough to verify our conclusion directly
from the equations of motion.  The particle Lagrangian can be
taken as $L = \tfrac{1}{2} \dot x \cdot \dot x$, which leads
directly to the following equation of motion:
\begin{equation}
    \ddot \ell = \frac{\varepsilon}{2} \frac{d\Omega}{d\ell} h_{\alpha\beta} u^\alpha
    u^\beta.
\end{equation}
Now, assuming that our particle is very close to $\ell = 0$, we
can expand $d\Omega/d\ell$ to first order in $\ell$ and
approximate $h_{\alpha\beta} u^\alpha u^\beta \sim -1$.  In both
of the relevant cases $\varepsilon\Lambda \lessgtr 0$, we obtain
\begin{equation}
    \ddot\ell = \frac{2\Lambda}{n(n-1)} \ell + \mathcal{O}(\ell^2)
    = (\text{sgn}\,\Lambda) \, \omega^2 \ell + \mathcal{O}(\ell^2).
\end{equation}
The obvious stability condition from this expression is $\Lambda <
0$, which matches the above result precisely.  Perhaps not
surprisingly, the frequency of oscillation about $\ell = 0$ is the
same frequency found in the warp factor.

We finish by noting that this metric \ansatz~easily lends itself
to toy models of spherically-symmetric astrophysical situations.
For example, suppose that we believed that there was a ---
suitably tiny --- cosmological constant $\Lambda_4$ permeating the
immediate vicinity of the Sun.  We could construct a 5-dimensional
description by taking $q_{\alpha\beta}$ to be the 4-metric around
a $\Lambda_4$-black hole:
\begin{subequations}
\begin{eqnarray}
    q_{\alpha\beta} dy^\alpha dy^\beta & = & -f(r)\,dt^2 +
    \frac{dr^2}{f(r)} + r^2 \, d\Omega_2^2, \\
    f(r) & = & 1 - \frac{2M}{r} - \frac{1}{3} \Lambda_4 r^2,
\end{eqnarray}
\end{subequations}
where $d\Omega_2^2$ is the metric on the unit 2-sphere.  Using
equation (\ref{warp Ricci}), we see that $\Lambda_4$ is related to
the 5-dimensional cosmological constant by
\begin{equation}
    \Lambda_4 = \tfrac{1}{3} A^2 \Lambda.
\end{equation}
Interestingly, $\Lambda_4$ has the same sign as $\Lambda$.  So in
this scenario, test particle like comets, asteroids and planets
will have a stable equilibrium about $\ell = 0$ if $\Lambda_4 <
0$; that is, the vacuum energy has negative density.  If
$\Lambda_4$ is positive, as suggested by recent cosmological
observations, the $\Sigma_0$ 4-manifold will be gravitationally
repulsive to test particles.  Hence, if a particle at $\ell = 0$
were to acquire a small extra dimensional velocity --- perhaps by
emitting gravitational radiation into the bulk --- there would be
no guarantee that it would return to our ``native'' spacetime.

\subsection{The Liu-Mashhoon-Wesson metric}\label{sec:lmw}

In \cite{Sea03a}, test particle trajectories in the following
5-geometry were considered:
\begin{subequations}\label{5-metric}
\begin{eqnarray}\label{5D Metric 2}
    \ds{M} & = & - b^2(t,\ell) dt^2 + a^2(t,\ell) d\sigma_k^2
    + d\ell^2, \\
    a^2(t,\ell) & \equiv & ( t^2 + k ) \ell^2 + \frac{{\mathcal K}}{t^2 + k}, \\
    b(t,\ell) & \equiv & \frac{ [( t^2 + k )^2 \ell^2 - {\mathcal K} ] }
    {(t^2 + k)^{3/2} [( t^2 + k )^2 \ell^2 + {\mathcal K} ]^{1/2} }, \\
    d\sigma_k^2 & \equiv & \frac{dr^2}{1-kr^2} + r^2 \, d\Omega_2^2.
\end{eqnarray}
\end{subequations}
Here, $\mathcal{K}$ is an integration constant and $k=0,\pm 1$.
This line element is a special case of one originally discovered
by Liu \& Mashhoon \cite{Liu95} and subsequently re-discovered by
Liu \& Wesson \cite{Liu01}, and is a solution of the 5-dimensional
vacuum field equations $\hat{R}_{AB} = 0$.  It is interesting
because the line element on each of the $\Sigma_\ell$
hypersurfaces is of the cosmological Robertson-Walker form.
However, this 5-geometry has recently been shown to be isometric
the 5-dimensional topological Schwarzschild solution
\cite{Sea03b,Sea03c}.

It is easy to confirm that $\ell = 0$ is a totally geodesic
4-surface in the geometry (\ref{5-metric}) with line element:
\begin{equation}\label{4d metric}
    \ds{\Sigma_0} = \frac{{\mathcal K}}{t^2+k} \left[ - \frac{dt^2}{(t^2 + k)^2}
    + d\sigma_k^2 \right],
\end{equation}
which can be shown to be isometric to a radiation-dominated
cosmology.  Note that in order to have a sensible solution, we
need to ensure that
\begin{equation}\label{kappa inequal}
    \frac{{\mathcal K}}{t^2+k} > 0
\end{equation}
by judiciously choosing $k$ and restricting the range of $t$.
Now, the tangent vector to a comoving geodesic path on $\Sigma_0$
is
\begin{equation}
    U^\alpha \di_\alpha = (t^2+k) \sqrt{\frac{t^2+k}{\mathcal{K}}}
    \di_t.
\end{equation}
The gravitational densities measured by such an observer are
easily calculated from the basic definitions (\ref{density defs}):
\begin{equation}
    \rho_g^{(5)} = 0, \quad \rho_g^{(4)} =
    \frac{3(t^2+k)^2}{\mathcal{K}}.
\end{equation}
Since the bulk is devoid of matter in this case, the stability of
the comoving trajectory on $\Sigma_0$ demands $\rho_g^{(4)} < 0$.
In other words, the $\ell = 0$ hypersurface will be
gravitationally attractive only if the strong energy condition is
\emph{violated} on $\Sigma_0$; i.e., if $\mathcal{K} < 0$.  Notice
that in order to have a negative value of the integration constant
$\mathcal{K}$, the inequality (\ref{kappa inequal}) implies that
$k = -1$ and we restrict $t \in (-1,1)$.

This conclusion is odd enough to warrant direct verification from
the higher-dimensional geodesic equation. The 5-dimensional
Lagrangian for comoving trajectories is
\begin{equation}
    L = \tfrac{1}{2} \left[ -b^2(t,\ell) \dot{t}^2 + \dot{\ell}^2
    \right].
\end{equation}
We can obtain an equation for $\ddot{\ell}$ by extremizing the
action, which yields
\begin{equation}\label{ell accn}
    \ddot\ell = -\frac{1}{2} \dot t^2 \frac{\di}{\di\ell} b^2(t,\ell) =
    \left( \frac{3\dot t^2}{t^2 + k} \right) \ell + O(\ell^3).
\end{equation}
We can use the solution for $U^\alpha$ above to approximate $\dot
t^2 \approx (t^2+k)^3/\mathcal{K}$ and write
\begin{equation}
    \ddot\ell = \frac{3(t^2+k)^2}{\mathcal{K}} \ell =
    \rho_g^{(4)} \ell.
\end{equation}
Here, we have omitted the higher-order terms from the equation of
motion.  This is what is expected from (\ref{linear extra}), and
confirms to us that the comoving trajectory with $\ell = 0$ is
stable only if $\rho_g^{(4)} < 0$.

\section{Summary and discussion: an energy condition for higher dimensions}\label{sec:summary}

In this paper, we have shown that confined particle trajectories
on totally geodesic $n$-surfaces embedded in $(n+1)$-dimensional
bulk manifolds are stable against small perturbations if
$\rho_g^{(n+1)} > \rho_g^{(n)}$.  Here, $\rho_g^{(n+1)}$ and
$\rho_g^{(n)}$ are the gravitational densities of $M$ and
$\Sigma_0$ as measured by observers travelling on the confined
trajectories, respectively.  We established this result using a
covariant decomposition of the higher-dimensional equation of
motion in Section \ref{sec:confinement}, the equation of geodesic
deviation in Section \ref{sec:deviation}, and the Raychaudhuri
equation in Section \ref{sec:raychaudhuri}.  In Section
\ref{sec:examples}, we gave several concrete examples of our
results involving warped product and Liu-Mashhoon-Wesson metrics.

We conclude by noting that the stability condition as formulated
above is only applicable to particular geodesic paths on
$\Sigma_0$.  For some applications, one might want to ensure that
\emph{all} the timelike trajectories through some region of
$\Sigma_0$ are stable against perturbations.  It is not difficult
to see how to generalize our previous results to satisfy this
stronger demand. Consider the following definition:
\begin{description}

\item[The Confinement Energy Condition:] Let $\Sigma_0$ be an
$n$-dimensional totally geodesic Lorentzian submanifold smoothly
embedded in $(n+1)$-dimensional bulk $M$.  Also, let ${}^0
R_{\alpha\beta}$ be the Ricci-tensor on $\Sigma_0$ and ${}^0
\hat{R}_{AB}$ be the Ricci-tensor on $M$, both evaluated at a
point $P \in \Sigma_0 \subset M$.  The confinement energy
condition at $P$ is
\begin{equation}
    ({}^0 \hat{R}_{AB} e^A_\alpha e^B_\beta - {}^0 R_{\alpha\beta})
    U^\alpha U^\beta > 0,
\end{equation}
where $U^\alpha$ is an arbitrary timelike vector tangent to
$\Sigma_0$.  If the confinement energy condition holds in some
neighbourhood $N[P] \subset \Sigma_0$ of $P$, then any test
particle travelling along a timelike trajectory on $\Sigma_0$
passing through $P$ will be stable against small perturbations
while in $N[P]$.

\end{description}
There are obviously significant similarities between this and the
familiar strong energy condition from 4-dimensional relativity,
and we note that it can be used to place conditions on the
densities and principle pressures associated with the Einstein
tensors of $M$ and $\Sigma_0$.  It is clear that for the examples
of Section \ref{sec:examples}, the confinement energy condition is
satisfied in the warped-product Einstein space situation if
$\Lambda < 0$, and in the Liu-Mashhoon-Wesson metric if the
4-dimensional strong energy condition is false on $\Sigma_0$.  We
have no doubt that it would be interesting to apply this condition
to other higher-dimensional situations with totally geodesic
submanifolds, but this is the subject for future work.

\begin{acknowledgments}
I would like to thank Paul Wesson and Tomas Liko for comments, and
NSERC and OGS for financial support.
\end{acknowledgments}

\bibliography{confinement}

\begin{thebibliography}{25}
\expandafter\ifx\csname natexlab\endcsname\relax\def\natexlab#1{#1}\fi
\expandafter\ifx\csname bibnamefont\endcsname\relax
  \def\bibnamefont#1{#1}\fi
\expandafter\ifx\csname bibfnamefont\endcsname\relax
  \def\bibfnamefont#1{#1}\fi
\expandafter\ifx\csname citenamefont\endcsname\relax
  \def\citenamefont#1{#1}\fi
\expandafter\ifx\csname url\endcsname\relax
  \def\url#1{\texttt{#1}}\fi
\expandafter\ifx\csname urlprefix\endcsname\relax\def\urlprefix{URL }\fi
\providecommand{\bibinfo}[2]{#2}
\providecommand{\eprint}[2][]{\url{#2}}

\bibitem[{\citenamefont{Arkani-Hamed et~al.}(1998)\citenamefont{Arkani-Hamed,
  Dimopoulos, and Dvali}}]{Ark98a}
\bibinfo{author}{\bibfnamefont{N.}~\bibnamefont{Arkani-Hamed}},
  \bibinfo{author}{\bibfnamefont{S.}~\bibnamefont{Dimopoulos}},
  \bibnamefont{and} \bibinfo{author}{\bibfnamefont{G.~R.} \bibnamefont{Dvali}},
  \bibinfo{journal}{Phys. Lett.} \textbf{\bibinfo{volume}{B429}},
  \bibinfo{pages}{263} (\bibinfo{year}{1998}),
  \eprint[http://arXiv.org/abs]{hep-ph/9803315}.

\bibitem[{\citenamefont{Arkani-Hamed et~al.}(1999)\citenamefont{Arkani-Hamed,
  Dimopoulos, and Dvali}}]{Ark98b}
\bibinfo{author}{\bibfnamefont{N.}~\bibnamefont{Arkani-Hamed}},
  \bibinfo{author}{\bibfnamefont{S.}~\bibnamefont{Dimopoulos}},
  \bibnamefont{and} \bibinfo{author}{\bibfnamefont{G.~R.} \bibnamefont{Dvali}},
  \bibinfo{journal}{Phys. Rev.} \textbf{\bibinfo{volume}{D59}},
  \bibinfo{pages}{086004} (\bibinfo{year}{1999}),
  \eprint[http://arXiv.org/abs]{hep-ph/9807344}.

\bibitem[{\citenamefont{Randall and Sundrum}(1999{\natexlab{a}})}]{Ran99a}
\bibinfo{author}{\bibfnamefont{L.}~\bibnamefont{Randall}} \bibnamefont{and}
  \bibinfo{author}{\bibfnamefont{R.}~\bibnamefont{Sundrum}},
  \bibinfo{journal}{Phys. Rev. Lett.} \textbf{\bibinfo{volume}{83}},
  \bibinfo{pages}{3370} (\bibinfo{year}{1999}{\natexlab{a}}),
  \eprint[http://arXiv.org/abs]{hep-ph/9905221}.

\bibitem[{\citenamefont{Randall and Sundrum}(1999{\natexlab{b}})}]{Ran99b}
\bibinfo{author}{\bibfnamefont{L.}~\bibnamefont{Randall}} \bibnamefont{and}
  \bibinfo{author}{\bibfnamefont{R.}~\bibnamefont{Sundrum}},
  \bibinfo{journal}{Phys. Rev. Lett.} \textbf{\bibinfo{volume}{83}},
  \bibinfo{pages}{4690} (\bibinfo{year}{1999}{\natexlab{b}}),
  \eprint[http://arXiv.org/abs]{hep-th/9906064}.

\bibitem[{\citenamefont{Khoury et~al.}(2001)\citenamefont{Khoury, Ovrut,
  Steinhardt, and Turok}}]{Kho01}
\bibinfo{author}{\bibfnamefont{J.}~\bibnamefont{Khoury}},
  \bibinfo{author}{\bibfnamefont{B.~A.} \bibnamefont{Ovrut}},
  \bibinfo{author}{\bibfnamefont{P.~J.} \bibnamefont{Steinhardt}},
  \bibnamefont{and} \bibinfo{author}{\bibfnamefont{N.}~\bibnamefont{Turok}},
  \bibinfo{journal}{Phys. Rev.} \textbf{\bibinfo{volume}{D64}},
  \bibinfo{pages}{123522} (\bibinfo{year}{2001}), \eprint{hep-th/0103239}.

\bibitem[{\citenamefont{Bucher}(2002)}]{Buc02}
\bibinfo{author}{\bibfnamefont{M.}~\bibnamefont{Bucher}},
  \bibinfo{journal}{Phys. Lett.} \textbf{\bibinfo{volume}{B530}},
  \bibinfo{pages}{1} (\bibinfo{year}{2002}).

\bibitem[{\citenamefont{Joseph}(1962)}]{Jos62}
\bibinfo{author}{\bibfnamefont{D.~W.} \bibnamefont{Joseph}},
  \bibinfo{journal}{Phys. Rev.} \textbf{\bibinfo{volume}{126}},
  \bibinfo{pages}{319} (\bibinfo{year}{1962}).

\bibitem[{\citenamefont{Akama}(1982)}]{Aka82}
\bibinfo{author}{\bibfnamefont{K.}~\bibnamefont{Akama}},
  \bibinfo{journal}{Lect. Notes Phys.} \textbf{\bibinfo{volume}{176}},
  \bibinfo{pages}{267} (\bibinfo{year}{1982}),
  \eprint[http://arXiv.org/abs]{hep-th/0001113}.

\bibitem[{\citenamefont{Rubakov and Shaposhnikov}(1983)}]{Rub83}
\bibinfo{author}{\bibfnamefont{V.~A.} \bibnamefont{Rubakov}} \bibnamefont{and}
  \bibinfo{author}{\bibfnamefont{M.~E.} \bibnamefont{Shaposhnikov}},
  \bibinfo{journal}{Phys. Lett.} \textbf{\bibinfo{volume}{B125}},
  \bibinfo{pages}{136} (\bibinfo{year}{1983}).

\bibitem[{\citenamefont{Visser}(1985)}]{Vis85}
\bibinfo{author}{\bibfnamefont{M.}~\bibnamefont{Visser}},
  \bibinfo{journal}{Phys. Lett.} \textbf{\bibinfo{volume}{B159}},
  \bibinfo{pages}{22} (\bibinfo{year}{1985}),
  \eprint[http://arXiv.org/abs]{hep-th/9910093}.

\bibitem[{\citenamefont{Gibbons and Wiltshire}(1987)}]{Gib87}
\bibinfo{author}{\bibfnamefont{G.~W.} \bibnamefont{Gibbons}} \bibnamefont{and}
  \bibinfo{author}{\bibfnamefont{D.~L.} \bibnamefont{Wiltshire}},
  \bibinfo{journal}{Nucl. Phys.} \textbf{\bibinfo{volume}{B287}},
  \bibinfo{pages}{717} (\bibinfo{year}{1987}), \eprint{hep-th/0109093}.

\bibitem[{\citenamefont{Wesson and Ponce~de Leon}(1992)}]{Wes92}
\bibinfo{author}{\bibfnamefont{P.~S.} \bibnamefont{Wesson}} \bibnamefont{and}
  \bibinfo{author}{\bibfnamefont{J.}~\bibnamefont{Ponce~de Leon}},
  \bibinfo{journal}{J. Math. Phys.} \textbf{\bibinfo{volume}{33}},
  \bibinfo{pages}{3883} (\bibinfo{year}{1992}).

\bibitem[{\citenamefont{Horava and Witten}(1996{\natexlab{a}})}]{Hor96a}
\bibinfo{author}{\bibfnamefont{P.}~\bibnamefont{Horava}} \bibnamefont{and}
  \bibinfo{author}{\bibfnamefont{E.}~\bibnamefont{Witten}},
  \bibinfo{journal}{Nuclear Physics} \textbf{\bibinfo{volume}{B460}},
  \bibinfo{pages}{506} (\bibinfo{year}{1996}{\natexlab{a}}),
  \eprint{hep-th/9510209}.

\bibitem[{\citenamefont{Horava and Witten}(1996{\natexlab{b}})}]{Hor96b}
\bibinfo{author}{\bibfnamefont{P.}~\bibnamefont{Horava}} \bibnamefont{and}
  \bibinfo{author}{\bibfnamefont{E.}~\bibnamefont{Witten}},
  \bibinfo{journal}{Nucl. Phys.} \textbf{\bibinfo{volume}{B475}},
  \bibinfo{pages}{94} (\bibinfo{year}{1996}{\natexlab{b}}),
  \eprint[http://arXiv.org/abs]{hep-th/9603142}.

\bibitem[{\citenamefont{Seahra and Wesson}(2003{\natexlab{a}})}]{Sea03a}
\bibinfo{author}{\bibfnamefont{S.~S.} \bibnamefont{Seahra}} \bibnamefont{and}
  \bibinfo{author}{\bibfnamefont{P.~S.} \bibnamefont{Wesson}},
  \bibinfo{journal}{Classical \& Quantum Gravity}
  \textbf{\bibinfo{volume}{20}}, \bibinfo{pages}{1321}
  (\bibinfo{year}{2003}{\natexlab{a}}), \eprint{gr-qc/0302015}.

\bibitem[{\citenamefont{DeWolfe et~al.}(2000)\citenamefont{DeWolfe, Freedman,
  Gubser, and Karch}}]{DeW99}
\bibinfo{author}{\bibfnamefont{O.}~\bibnamefont{DeWolfe}},
  \bibinfo{author}{\bibfnamefont{D.~Z.} \bibnamefont{Freedman}},
  \bibinfo{author}{\bibfnamefont{S.~S.} \bibnamefont{Gubser}},
  \bibnamefont{and} \bibinfo{author}{\bibfnamefont{A.}~\bibnamefont{Karch}},
  \bibinfo{journal}{Phys. Rev.} \textbf{\bibinfo{volume}{D62}},
  \bibinfo{pages}{046008} (\bibinfo{year}{2000}),
  \eprint[http://arXiv.org/abs]{hep-th/9909134}.

\bibitem[{\citenamefont{Csaki et~al.}(2000)\citenamefont{Csaki, Erlich,
  Hollowood, and Shirman}}]{Csa00}
\bibinfo{author}{\bibfnamefont{C.}~\bibnamefont{Csaki}},
  \bibinfo{author}{\bibfnamefont{J.}~\bibnamefont{Erlich}},
  \bibinfo{author}{\bibfnamefont{T.~J.} \bibnamefont{Hollowood}},
  \bibnamefont{and} \bibinfo{author}{\bibfnamefont{Y.}~\bibnamefont{Shirman}},
  \bibinfo{journal}{Nucl. Phys.} \textbf{\bibinfo{volume}{B581}},
  \bibinfo{pages}{309} (\bibinfo{year}{2000}),
  \eprint[http://arXiv.org/abs]{hep-th/0001033}.

\bibitem[{\citenamefont{Seahra}(2002)}]{Sea02}
\bibinfo{author}{\bibfnamefont{S.~S.} \bibnamefont{Seahra}},
  \bibinfo{journal}{Physical Review} \textbf{\bibinfo{volume}{D65}},
  \bibinfo{pages}{124004} (\bibinfo{year}{2002}),
  \eprint[http://arXiv.org/abs]{gr-qc/0204032}.

\bibitem[{\citenamefont{Seahra}(2003)}]{Sea03b}
\bibinfo{author}{\bibfnamefont{S.~S.} \bibnamefont{Seahra}}, Ph.D. thesis,
  \bibinfo{school}{University of Waterloo} (\bibinfo{year}{2003}).

\bibitem[{\citenamefont{Liu and Mashhoon}(1995)}]{Liu95}
\bibinfo{author}{\bibfnamefont{H.}~\bibnamefont{Liu}} \bibnamefont{and}
  \bibinfo{author}{\bibfnamefont{B.}~\bibnamefont{Mashhoon}},
  \bibinfo{journal}{Annalen der Physik} \textbf{\bibinfo{volume}{4}},
  \bibinfo{pages}{565} (\bibinfo{year}{1995}).

\bibitem[{\citenamefont{Liu and Wesson}(2001)}]{Liu01}
\bibinfo{author}{\bibfnamefont{H.}~\bibnamefont{Liu}} \bibnamefont{and}
  \bibinfo{author}{\bibfnamefont{P.~S.} \bibnamefont{Wesson}},
  \bibinfo{journal}{Astrophysical Journal} \textbf{\bibinfo{volume}{562}},
  \bibinfo{pages}{1} (\bibinfo{year}{2001}), \eprint{gr-qc/0107093}.

\bibitem[{\citenamefont{Mashhoon et~al.}(1994)\citenamefont{Mashhoon, Liu, and
  Wesson}}]{Mas94}
\bibinfo{author}{\bibfnamefont{B.}~\bibnamefont{Mashhoon}},
  \bibinfo{author}{\bibfnamefont{H.}~\bibnamefont{Liu}}, \bibnamefont{and}
  \bibinfo{author}{\bibfnamefont{P.}~\bibnamefont{Wesson}},
  \bibinfo{journal}{Physics Letters} \textbf{\bibinfo{volume}{B331}},
  \bibinfo{pages}{305} (\bibinfo{year}{1994}).

\bibitem[{\citenamefont{Wesson}(1999)}]{Wes99}
\bibinfo{author}{\bibfnamefont{P.~S.} \bibnamefont{Wesson}},
  \emph{\bibinfo{title}{Space-Time-Matter}} (\bibinfo{publisher}{World
  Scientific}, \bibinfo{address}{Singapore}, \bibinfo{year}{1999}).

\bibitem[{\citenamefont{Frolov et~al.}(2003)\citenamefont{Frolov, Snajdr, and
  Stojkovic}}]{Fro03}
\bibinfo{author}{\bibfnamefont{V.}~\bibnamefont{Frolov}},
  \bibinfo{author}{\bibfnamefont{M.}~\bibnamefont{Snajdr}}, \bibnamefont{and}
  \bibinfo{author}{\bibfnamefont{D.}~\bibnamefont{Stojkovic}}
  (\bibinfo{year}{2003}), \eprint{gr-qc/0304083}.

\bibitem[{\citenamefont{Seahra and Wesson}(2003{\natexlab{b}})}]{Sea03c}
\bibinfo{author}{\bibfnamefont{S.~S.} \bibnamefont{Seahra}} \bibnamefont{and}
  \bibinfo{author}{\bibfnamefont{P.~S.} \bibnamefont{Wesson}}
  (\bibinfo{year}{2003}{\natexlab{b}}), \bibinfo{note}{in press in the Journal
  of Mathematical Physics}, \eprint{gr-qc/0309006}.

\end{thebibliography}

\end{document}